\documentclass{article}
\usepackage{graphicx} 
\usepackage{amsmath}
\usepackage{amsfonts}
\usepackage{bm}
\usepackage{braket}
\usepackage{ragged2e}
\usepackage{float}
\usepackage{subcaption}
\usepackage{xcolor}
\usepackage{booktabs}  
\usepackage{makecell} 
\usepackage{float}

\usepackage [left = 2.5 cm, right = 2.5 cm, top = 2.5 cm, bottom = 2.5 cm, bindingoffset = 0 cm]{geometry}

\emergencystretch=\maxdimen
\title{Solving the MAX-$k$-CUT problem for hypergraphs with quantum computing}

\author{
	\textbf{V. S. Yarmus}\thanks{e-mail: vitalii.yarmus.res@gmail.com} \quad 
	\textbf{Kh. P. Gnatenko}\thanks{e-mail: khrystyna.gnatenko@gmail.com} \\[2ex]
	\small Ivan Franko National University of Lviv,\\
	\small Professor Ivan Vakarchuk Department for Theoretical Physics,\\
	\small 12, Drahomanov St., Lviv, UA-79005, Ukraine
}

\begin{document}
	
	\maketitle
	
\begin{abstract}
We investigate the MAX-$k$-CUT problem on hypergraphs by considering their representation as spin systems with Ising interactions. The corresponding cost Hamiltonian for MAX-$k$-CUT problem on hypergraphs is reduced to an Ising Hamiltonian with pairwise interactions. This reduction simplifies the solution of the MAX-$k$-CUT problem using quantum computing. As illustrative examples, a 3-uniform tight-star hypergraph and a sunflower hypergraph are considered. The corresponding Hamiltonians are constructed, and quantum protocols for the MAX-$k$-CUT problems are presented.
The MAX-3-CUT and MAX-4-CUT instances are investigated for these hypergraphs by varying the number of Quantum Approximate Optimization Algorithm (QAOA) layers from (p=2) to (p=15) with quantum computing on IBM's Qiskit AerSimulator. 
\end{abstract}

\maketitle

\section{Introduction}

Hybrid Quantum-Classical algorithms based on the variational principle have attracted considerable attention due to their potential for solving NP-hard problems on noisy intermediate-scale quantum (NISQ) computers. Such algorithms represent a first step in leveraging quantum advantage to solve practical problems on existing quantum hardware.

A prominent example is the Quantum Approximate Optimization Algorithm (QAOA) \cite{farhi2014quantum}, which finds approximate solutions to combinatorial optimization problems. This algorithm has received significant attention recently due to its ability to find reasonable approximate solutions to complex problems, such as Maximum Cut (MAX-CUT), Maximum Independent Set (MIS), Binary Paint Shop Problem (BPSP), Binary Linear Least Squares (BLLS), and other problems that can be reduced to the quadratic unconstrained binary optimization (QUBO) problem \cite{moussa2022unsupervised, blekos2024review}. In this paper, we extend the application of QAOA to solve the MAX-CUT and MAX-$k$-CUT problems specifically for hypergraphs.

Graph theory has found significant practical applications in the modern world. Graphs are collections of objects with connections between them, where objects are represented by vertices and connections by edges. Graphs are actively used in telecommunication networks, machine learning algorithms, social networks, and many other areas \cite{majeed2020graph}. Since many graph-related problems are NP-hard, quantum or hybrid algorithms may offer computational advantages compared to purely classical approaches.

Hypergraphs are a generalization of graphs where edges, often called hyperedges, can connect any number of vertices. Hypergraphs provide a natural representation for systems where multiple elements interact simultaneously. Many systems with higher-order interactions in biology, ecology, neurology, and social sciences can be represented using hypergraphs \cite{antelmi2023survey}. Despite their convenience and modeling advantages, hypergraphs are less studied than ordinary graphs. However, interest in this topic is currently growing, partly due to research showing their necessity in describing systems with nonlinear interactions \cite{battiston2020networks}.

Currently, researchers are actively developing techniques to represent hypergraph structures within quantum states \cite{rossi2013quantum, salem2025quantum}. This progress has enabled quantum algorithms for challenges like hypergraph sparsification, the Minimum Cut problem, and the Minimum $s$-$t$ Cut \cite{liu2025quantum}. Furthermore, QAOA has been adapted for problems such as Max-q-XORSAT \cite{basso2022performance} and 3-SAT \cite{zhang2024grover}.

The remainder of this paper is organized as follows: Section 2 reviews the QAOA framework and its application to the MAX-CUT problem, including extensions for the MAX-$k$-CUT problem on graphs. Section 3 applies QAOA to these problems on hypergraphs. The cost Hamiltonian is reduced to pairwise interactions, thereby simplifying the problem and enabling the reuse of standard graph-based QAOA properties. In the section we also present results of studies MAX-$k$-CUT problem for 3-uniform tight star hypergraph and a sunflower hypergraph with quantum computing. Conclusions are presented in section 4.

\section{Solving MAX-CUT and MAX-$k$-CUT problems on graphs}

The maximum cut (MAX-CUT) problem is a well-known NP-hard optimization problem. The goal is to partition the vertices of a graph into two complementary sets such that the number of edges between vertices in different sets is maximized \cite{blekos2024review}. This can also be defined for weighted graphs, where the sum of the weights of edges between sets is maximized. Since solving this problem on classical computers is exponentially complex, much research focuses on  solutions with quantum computing.

The Quantum Approximate Optimization Algorithm (QAOA) was developed to find approximate solutions to combinatorial optimization problems \cite{farhi2014quantum}. In principle, this algorithm can solve any problem that can be formulated as a Quadratic Unconstrained Binary Optimization (QUBO) problem, which is mathematically equivalent to finding the ground state of the Ising model \cite{lucas2014ising, blekos2024review}.

QAOA optimizes the expectation value of a cost Hamiltonian $\hat{H}$ evaluated on a parameterized quantum state $\ket{\psi_p(\vec{\gamma}, \vec{\beta})}$. For the MAX-CUT problem on a weighted graph $G=(V,E)$, the cost Hamiltonian is defined as:
\begin{equation}
    \hat{H}=\frac{1}{2}\sum_{\left<i,j\right> \in E}{\omega_{ij}(1 - \sigma^z_i \sigma^z_j)}.
\end{equation}
The Hamiltonian corresponds to an Ising model. Coupling constants $\omega_{ij}$ represent the weight of the edge connecting vertices $i$ and $j$. The quantum state $\ket{\psi_p(\vec{\gamma}, \vec{\beta})}$ is created by alternately applying the cost $\hat{U}_C(\gamma_r)$ and mixer $\hat{U}_M(\beta_r)$ operators for $p$ iterations:
\begin{equation}
\ket{\psi_p(\vec{\gamma}, \vec{\beta})}=\prod_{r=p}^{1}\hat{U}_M(\beta_r)\hat{U}_C(\gamma_{r})\ket{s}.
\end{equation}
The layer unitaries are defined as:
\begin{eqnarray}
    \hat{U}_C(\gamma_r)=e^{-i \gamma_r \hat{H}} = \prod_{\left<i,j\right> \in E} Rzz_{ij}(-\omega_{ij}\gamma_r),
\end{eqnarray}
\begin{eqnarray}
    \hat{U}_M(\beta_r)=e^{-i \beta_r \hat{H}_M} = \prod_{i=1}^{n} Rx_i(2\beta_r),
\end{eqnarray}
 where $\hat{H}_M$ is called a mixer Hamiltonian and defined as:
\begin{eqnarray}
    \hat{H}_M = \sum_{i = 1}^n{\sigma^x_i},
\end{eqnarray}
 and the initial state is defined as:
\begin{eqnarray}
\ket{s} = \ket{+}^{\otimes n},
\end{eqnarray}
\begin{eqnarray}
\ket{+}=\frac{1}{\sqrt{2}}(\ket{0}+\ket{1}),
\end{eqnarray}
 where $n$ is the number of qubits (spins). Operators $Rx_i(2\beta_r)$, $Rzz_{ij}(-\omega_{ij}\gamma_r)$ are given by:
\begin{eqnarray}
 Rx_i(2\beta_r)=e^{-i\beta_r \sigma^x_i},
\end{eqnarray}
\begin{eqnarray}
Rzz_{ij}(-\omega_{ij}\gamma_r)=e^{i\omega_{ij}\frac{\gamma_r}{2}\sigma^z_i\sigma^z_j},
\end{eqnarray}
 where $\sigma^x_i$, $\sigma^z_i$ are Pauli matrices corresponding to qubit $q[i]$ (spin $i$).
The expectation value is defined as:
\begin{equation}
\bra{\psi_p(\vec{\gamma}, \vec{\beta})}\hat{H}\ket{\psi_p(\vec{\gamma}, \vec{\beta})}=F(\vec{\gamma}, \vec{\beta})
\end{equation}

A classical optimizer searches for  $\vec{\gamma}$ and $\vec{\beta}$ that maximize $F(\vec{\gamma}, \vec{\beta})$. The accuracy typically increases with the number of iterations $p$. We can calculate the approximation ratio $\alpha$ as:
\begin{equation}
\alpha=\frac{F(\vec{\gamma}^*, \vec{\beta}^*)}{C_{max}},
\end{equation}
where $\vec{\gamma}^*$ and $\vec{\beta}^*$ are optimal parameters and $C_{max}$ is the maximum cut value.

The MAX-CUT problem can be generalized to divide vertices into $k$ complementary sets, known as the MAX-$k$-CUT problem. Modifications of QAOA have been proposed to solve this \cite{hadfield2019quantum, fuchs2021efficient, bako2025prog} using three different encoding types.
Among them is qudit encoding: one qudit represents one vertex; set membership is encoded in the qudit basis states. One-hot encoding \cite{hadfield2019quantum, bako2025prog}: $k$ qubits represent one vertex; belonging to a set is encoded by the state $\ket{1}$ in a single qubit, while the others are $\ket{0}$.
Binary encoding \cite{fuchs2021efficient}: $\lceil\log_2(k)\rceil$ qubits represent one vertex; set membership is encoded in binary (e.g., for $k=4$, states are $\ket{00}, \ket{01}, \ket{10}, \ket{11}$).

Although binary encoding is more efficient, Hamiltonian construction is challenging, so this paper utilizes one-hot encoding. The cost Hamiltonian for it is given by:
\begin{equation}
\label{eq:graph_max_cut_hamiltonian}
\hat{H}_C=\sum_{\langle i,j \rangle \in E}{\omega_{ij}\sum_{a=1}^k {\sigma^z_{i,a} \sigma^z_{j,a}}}.
\end{equation}
where $k$ is the number of complementary sets and $\sigma^z_{i,a}$ is the Pauli matrix acting on the $a$-th qubit of vertex $i$. 

Under this formulation, cut edges contribute less energy ($k-4$) than uncut
edges ($k$). Consequently, finding the ground state of \eqref{eq:graph_max_cut_hamiltonian} (i.e., minimizing the expectation value $F(\vec{\gamma}, \vec{\beta})$) directly maximizes the total weight of cut edges.
Incorporating the feasibility constraint into the mixer Hamiltonian yields the $XY$-mixer \cite{hadfield2019quantum, fuchs2021efficient, wang2020xy}. In this paper the $XY$-mixer with ring topology is used:
\begin{equation}
    \hat{H}_{XY}^{ring} = \sum_{v \in V}\sum_{a=1}^{k}{( \sigma^x_{v,a}\sigma^x_{v,a+1}+\sigma^y_{v,a}\sigma^y_{v,a+1})}.
\end{equation}
where $\sigma^x_{v,a}$ and $\sigma^y_{v,a}$ are Pauli matrices acting on the $a$-th qubit of vertex $v$. 

Since this Hamiltonian contains non-commutative terms, a first-order approximation called a parity-partitioned mixer is used in practice:
\begin{equation}
    \hat{U}_{XY, v}(\beta)=\hat{U}_{last,v}(\beta)\hat{U}_{even,v}(\beta)\hat{U}_{odd,v}(\beta),
\end{equation}
where:
\begin{eqnarray}
\hat{U}_{odd,v}(\beta)=\prod_{a\text{ odd}, a \neq k}^k e^{-i\beta(\sigma^x_{v,a}\sigma^x_{v,a+1}+\sigma^y_{v,a}\sigma^y_{v,a+1})},\\
\hat{U}_{even,v}(\beta)=\prod_{a\text{ even}}^k e^{-i\beta(\sigma^x_{v,a}\sigma^x_{v,a+1}+\sigma^y_{v,a}\sigma^y_{v,a+1})},\\
\hat{U}_{last,v}(\beta)=\left\{
    \begin{array}{rcl}
    e^{-i\beta(\sigma^x_{v,k}\sigma^x_{v,1}+ \sigma^y_{v,k}\sigma^y_{v,1})} &  & k\text{ odd} \\
    \hat{I} &  & k\text{ even}
    \end{array}\right. ,
\end{eqnarray}
with periodic boundary conditions applied such that $k+1 \equiv 1$.

The initial state for one-hot encoding should be a superposition of all feasible vertex states: $\ket{\phi_0}=\ket{W_k}^{\otimes|V|}$, where:
\begin{eqnarray}
    \ket{W_k}=\frac{1}{\sqrt{k}}(\ket{100\dots00}+\ket{010\dots00}+\dots\nonumber\\+\ket{000\dots01}).
\end{eqnarray}
An efficient algorithm for the $W$ state is presented in \cite{cruz2019efficient}.

Because the one-hot encoding assigns an energy of $k$ to uncut edges and $k-4$ to cut edges, the energy spectrum is shifted by a constant offset. Let $E_{\min}$ be the ground state energy of the cost Hamiltonian \eqref{eq:graph_max_cut_hamiltonian}. We can directly map  $F(\vec{\gamma}^*, \vec{\beta}^*)$ and $E_{\min}$ to their corresponding cut weights:
\begin{eqnarray}
C_{\text{exp}} = \frac{k W_{\text{total}} - F(\vec{\gamma}^*,\vec{\beta}^*)}{4},\\ C_{\max} = \frac{k W_{\text{total}} - E_{\min}}{4},
\end{eqnarray}
where $W_{\text{total}}$ is the total weight of the graph's edges. 

The approximation ratio $\alpha$ for the MAX-$k$-CUT problem can thus be formulated in the following form:
\begin{equation}
\alpha = \frac{C_{\text{exp}}}{C_{\max}} = \frac{k W_{\text{total}} - F(\vec{\gamma}^*, \vec{\beta}^*)}{k W_{\text{total}} - E_{\min}}.
\end{equation}
 
\section{Solving MAX-CUT and MAX-$k$-CUT on hypergraphs}

Generalizing the MAX-CUT problem for hypergraphs requires a construction of Hamiltonian whose ground state provides the solution. 
 For solving MAX-CUT problem on hypergraphs, we consider replacing each hyperedge with a complete graph (a graph in which every pair of vertices is connected by an edge). This process is called the 2-section of the hypergraph. As a result, the Hamiltonian can be written as follows:

\begin{equation}
\label{eq:hypergraph_max_cut_ham}
    \hat{H}_C=\sum_{e \in E}{\omega_e \sum_{i,j \in e; i<j}{\sigma^z_i \sigma^z_j}} ,
\end{equation}
where $\omega_e$ is the weight of the hyperedge and $E$ is the set of hyperedges.

The Hamiltonian corresponds to an Ising model. It utilizes only double interactions, simplifying the hypergraph to a standard graph representation. When summing all terms with the same $\sigma^z_i \sigma^z_j$, we obtain a Hamiltonian corresponding to graph. In the worst case, the total number of double interactions equals:
\begin{equation}
    \binom{|V|}{2}=\frac{|V|(|V|-1)}{2},
\end{equation}
where $|V|$ is the number of vertices in the hypergraph.

To show that the smallest eigenvalue of \eqref{eq:hypergraph_max_cut_ham} corresponds to a balanced cut of a hyperedge, consider a single hyperedge with $n$ vertices. It has $n(n-1)/2$ edges. Each term $\sigma^z_i\sigma^z_j$ gives $+1$ if the vertices are in the same set and $-1$ if they are in different sets. The eigenvalue is minimized when the number of pairs in different subsets is maximized. For $n$ vertices split into two sets of sizes $n_1$ and $n_2$, the number of interactions between the sets is maximized when $n_1 = n_2 = n/2$, as follows from the AM-GM inequality:

\begin{equation}
\frac{n_1+n_2}{2}\geq\sqrt{n_1 n_2}\Rightarrow \frac{n^2}{4}\geq n_1 n_2.
\end{equation}

Thus, the optimal cut into two subsets occurs when $n_1=\lceil n/2\rceil$ and $n_2=\lfloor n/2\rfloor$. This Hamiltonian strives for balanced cuts of each individual hyperedge, though it does not guarantee that each group will have an identical number of vertices globally.

The eigenvalue for a specific hyperedge can be calculated based on the number of vertices in each group. The number of double interactions between the different groups is $n_1 n_2$. Since these terms contribute $-1$, and the remaining $n(n-1)/2 - n_1 n_2$ terms contribute $+1$, the eigenvalue is 
$E_0=\omega_e(n(n-1)/2-2n_1 n_2)$.

The fact that we simplified a hypergraph to a graph allows us to reuse the one-hot encoding approach for hypergraphs. The Hamiltonian becomes:
\begin{equation}
    \hat{H}_P=\sum_{e \in E}{\omega_e \sum_{i,j \in e; i<j}{\sum_{a=1}^k\sigma^z_{i,a} \sigma^z_{j,a}}},
\end{equation}
where $k$ is the number of complementary sets.

In our implementation, we used static default parameters $\gamma=\pi/2$ and $\beta=\pi/4$ for each layer. The classical optimizer was COBYLA with $maxiter=2000$ and number of shots set to $10000$.
To calculate the number of qubits and CNOT gates required for the protocol, we can use Equation \eqref{eq:q-num-and-cx}. The solution will provide an upper bound because the problem Hamiltonian can be simplified for certain hypergraphs due to the duplicated $\sigma_i^z\sigma_j^z$ operators when the hyperedges overlap.

\begin{equation}
\begin{split}
& \#q = |V| k \\
& \#CX = CX_W + p(CX_{U_P} + CX_{U_{XY}})  \\
& CX_W = |V|(2k-2) \hfill \\
& CX_{U_P} = \sum_{e \in E}{\sum_{i,j \in e; i<j}{2k}} \\
& CX_{U_{XY}} = |V|4k,
\end{split}
\
\label{eq:q-num-and-cx}
\end{equation}
where $|V|$ -- number of vertices in hypergraph, $k$ - number of sets the hypergraph is divided into, $p$ -- number of layers and $E$ -- set of hyperedges.

\begin{figure}
    \centering
      \caption{A 3-uniform hypergraph which is a tight star of size 3.}
        
        \label{fig:3-uniform-tight-star}
        \vskip3mm\tabcolsep10pt
    \begin{minipage}{0.4\textwidth}
        \centering
        \includegraphics[width=1\textwidth]{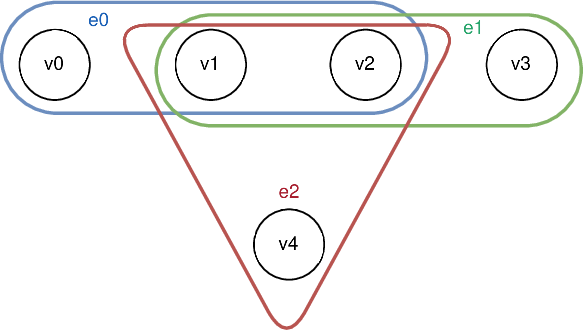}
    \end{minipage}
   \centering
     \caption{A sunflower hypergraph.}
        \label{fig:sunflower}
    \begin{minipage}{0.4\textwidth}
        \centering
    \includegraphics[width=1\textwidth]{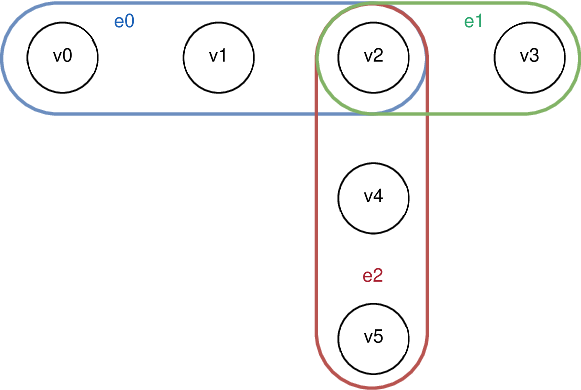}   
    \end{minipage}
\end{figure}

\begin{figure}[H]
    \centering
    \vskip3mm\tabcolsep10pt
    \caption{The quantum circuit of QAOA for MAX-$k$-CUT of the tight star hypergraph with $k = 3$ and $p = 1$.}
    \vskip3mm\tabcolsep15pt
    \label{fig:tight-star-cirquit}
    \includegraphics[width=0.95\textwidth]{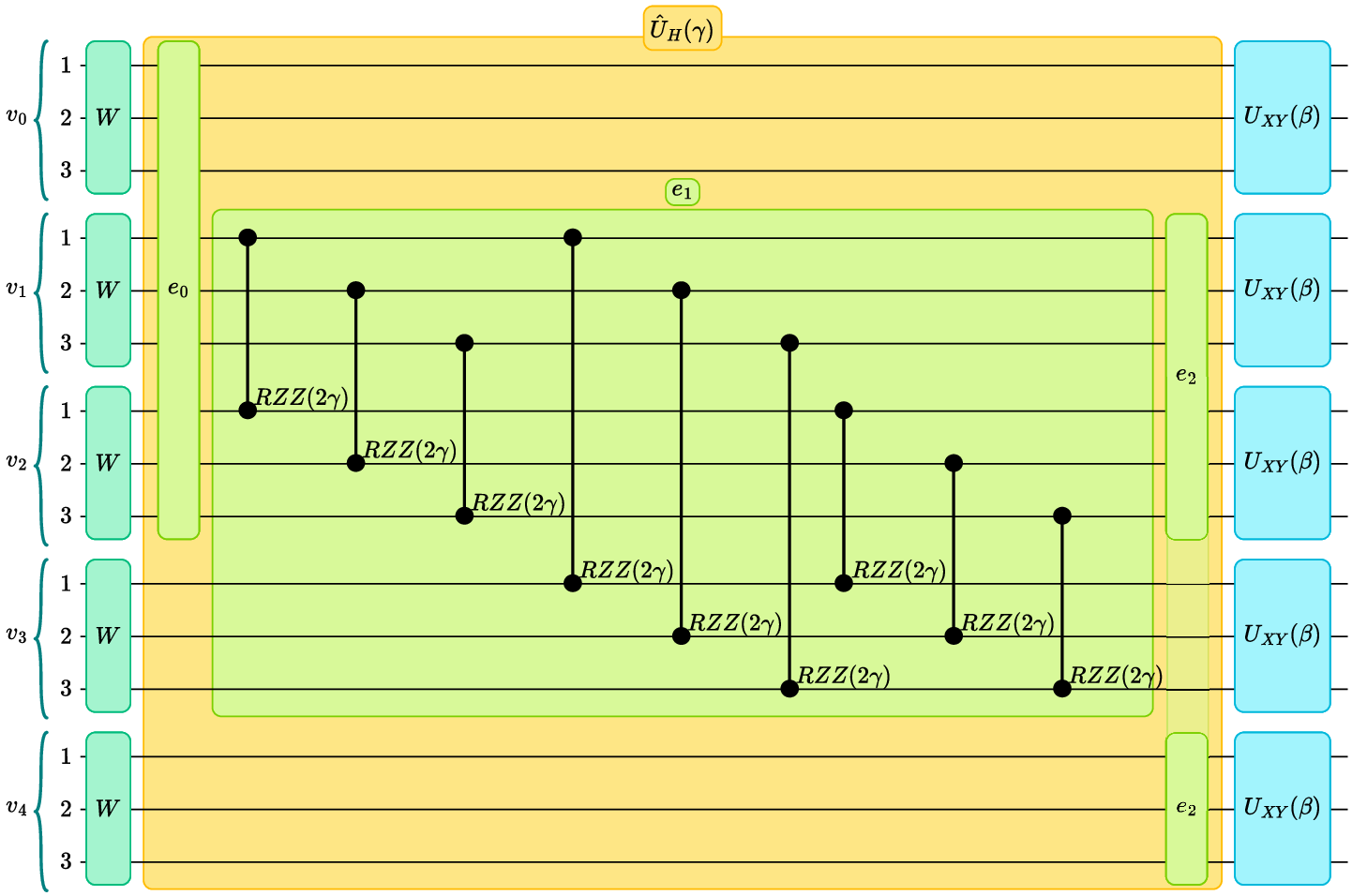}
\end{figure}

The first example is a 3-uniform tight star hypergraph shown in Fig. \ref{fig:3-uniform-tight-star}. The quantum circuit for $k=3$ and $p=1$ is depicted in Fig. \ref{fig:tight-star-cirquit}. Results for different numbers of layers are presented in Table \ref{tbl:3-uniform-tight-star}. As observed, the approximation ratio does not improve consistently with $p$. This is due to local minima in the search space where the classical optimizer may get stuck.

\begin{table}[H]
    \centering
    \vskip3mm\tabcolsep10pt
\caption{\normalsize  Performance metrics for tight star hypergraph cut into $k$ sets with the numbers of layers -- $p$, approximation ratio -- $\alpha$, CNOT counts -- $\#CX$ and number of qubits -- $\#q$. The results determine whether the mode of the distribution corresponds to a global optimum.}
    \label{tbl:3-uniform-tight-star}
     \vskip3mm\tabcolsep18pt
    {\scriptsize
    \begin{tabular}{cccllll}
        \toprule
        \textbf{k} & \textbf{\#q} & \textbf{\#CX} & \textbf{p=2} & \textbf{p=5} & \textbf{p=10} & \textbf{p=15} \\
        \midrule
        3 & 15 & $20+102p$ 
        & \makecell[l]{$\alpha=0.73$ \\ not optimal} 
        & \makecell[l]{$\alpha=0.75$ \\ optimal} 
        & \makecell[l]{$\alpha=0.82$ \\ optimal} 
        & \makecell[l]{$\alpha=0.78$ \\ not optimal} \\
        \midrule
        4 & 20 & $30+136p$ 
        & \makecell[l]{$\alpha=0.88$ \\ optimal} 
        & \makecell[l]{$\alpha=0.93$ \\ optimal} 
        & \makecell[l]{$\alpha=0.93$ \\ optimal} 
        & \makecell[l]{$\alpha=0.94$ \\ optimal} \\
        \bottomrule
    \end{tabular}
    }
    
\end{table}

\begin{figure}[H]
    \centering
    \vskip3mm\tabcolsep10pt
    \caption{The quantum circuit of QAOA for MAX-$k$-CUT of the sunflower hypergraph with $k = 3$ and $p = 1$.}
    \vskip3mm\tabcolsep15pt
    \label{fig:sunflower-cirquit}
    \includegraphics[width=0.95\textwidth]{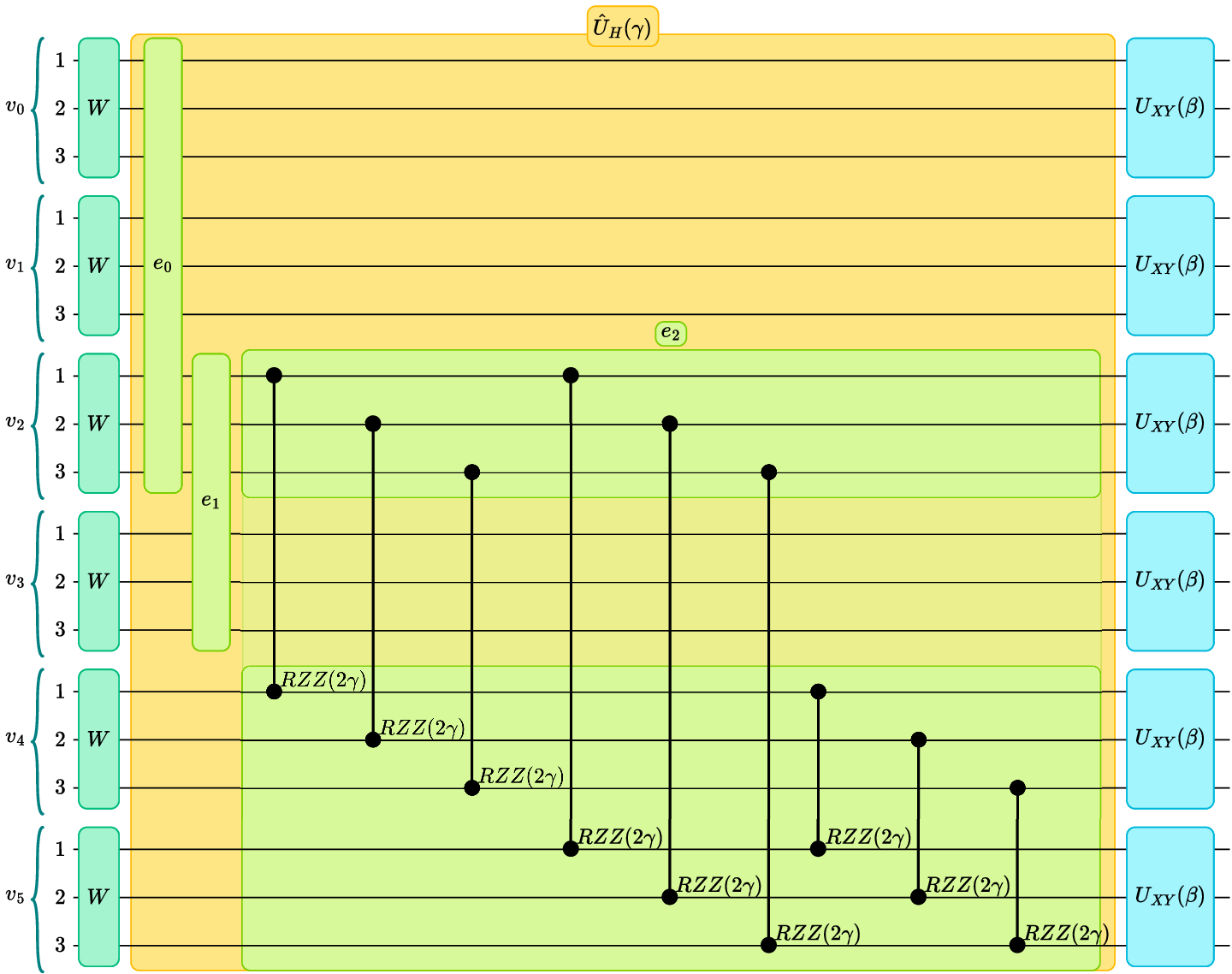}
\end{figure}

The second example is the sunflower hypergraph in Fig. \ref{fig:sunflower}, with the quantum circuit for $k=3$ and $p=1$ depicted in Fig. \ref{fig:sunflower-cirquit} and results in Table 2. 

\begin{table}[H]
\label{tbl:sunflower} 
\noindent
\caption{Performance metrics for sunflower hypergraph cut into $k$ sets with the number of layers -- $p$, approximation ratio -- $\alpha$, CNOT counts -- $\#CX$, and number of qubits -- $\#q$. The results determine whether the mode of the distribution corresponds to a global optimum.}
\vskip3mm
\setlength{\tabcolsep}{18pt}

\noindent{\footnotesize
\begin{tabular}{cccllll}
    \toprule
    \textbf{k} & \textbf{\#q} & \textbf{\#CX} &
    \textbf{p=2} & \textbf{p=5} & \textbf{p=10} & \textbf{p=15} \\
    \midrule
    3 & 18 & $24+114p$
    & \makecell[l]{$\alpha=0.67$ \\ not optimal}
    & \makecell[l]{$\alpha=0.67$ \\ not optimal}
    & \makecell[l]{$\alpha=0.86$ \\ optimal}
    & \makecell[l]{$\alpha=0.67$ \\ not optimal} \\
    \midrule
    4 & 24 & $36+152p$
    & \makecell[l]{$\alpha=0.75$ \\ not optimal}
    & \makecell[l]{$\alpha=0.75$ \\ not optimal}
    & \makecell[l]{$\alpha=0.75$ \\ not optimal}
    & \makecell[l]{$\alpha=0.75$ \\ not optimal} \\
    \bottomrule
\end{tabular}}
\end{table}

\section{Conclusions}

We have investigated the MAX-$k$-CUT problem on hypergraphs
by considering their 2-section and mapping the resulting graph to a spin system described by the Ising model. This approach enables the corresponding cost Hamiltonian to be formulated in terms of an Ising Hamiltonian with pairwise interactions, providing a convenient framework for studying hypergraph partitioning problems using quantum computing methods.

Numerical simulation results for the 3-uniform tight star and sunflower hypergraphs demonstrate that the global optimum can be achieved. However, simulations also show that the implementation is susceptible to local minima issues inherited from $XY$-QAOA \cite{wang2020xy}. This is expected as the hypergraph is simplified into a graph representation. This suggests that the 2-section of a hypergraph is an effective technique for solving specific hypergraph problems within quantum programming.

For both cases, a star hypergraph and a sunflower hypergraph, the MAX-3-CUT and MAX-4-CUT optimization problems are analyzed using QAOA circuits with depths ranging from p=2 to p=15. The simulation results highlight the algorithm's potential for solving combinatorial optimization problems defined on hypergraphs.

\vskip3mm \textit{Acknowledgement.} The work of Kh. P. Gnatenko was supported  by Project 2025.07/0326 (No. 0126U002943) from the National Research Foundation of Ukraine.

\end{document}